\documentclass[prd,showpacs,showkeys,nofootinbib,twocolumn]{revtex4-1}

\usepackage{graphicx}
\usepackage{amsmath,amsfonts,amssymb,amsxtra,latexsym}

\usepackage{hyperref}
\hypersetup{
    colorlinks=true,
    linkcolor=red,
    citecolor=blue,      
    urlcolor=blue,
}

\usepackage[utf8]{inputenc}

\begin{document}

\title{Plane Fronted Limit of Spherical Electromagnetic and Gravitational Waves}

\author{Peter A. Hogan}
\email{peter.hogan@ucd.ie}
\affiliation{School of Physics, University College Dublin, Belfield, Dublin 4, Ireland} 

\author{Dirk Puetzfeld}
\email{dirk.puetzfeld@zarm.uni-bremen.de}
\homepage{http://puetzfeld.org}
\affiliation{University of Bremen, Center of Applied Space Technology and Microgravity (ZARM), 28359 Bremen, Germany} 

\date{ \today}

\begin{abstract}
We demonstrate how plane fronted waves with colliding wave fronts are the asymptotic limit of spherical electromagnetic and gravitational waves. In the case of the electromagnetic waves we utilize Bateman's representation of radiative solutions of Maxwell's vacuum field equations. The gravitational case involves a novel form of the radiative Robinson--Trautman solutions of Einstein's vacuum field equations.
\end{abstract}


\maketitle


\section{Introduction} \label{sec:introduction}

The Robinson--Trautman \cite{Robinson:Trautman:1960,Robinson:Trautman:1962} spherical, noncolliding, gravitational waves propagating in a vacuum have been shown to have an asymptotic limit of pp--waves \cite{Robinson:Trautman:1960} (plane fronted waves with noncolliding wave fronts). The Robinson--Trautman waves have an isolated source and so one expects the waves to collide in general depending upon how the source is moving. Consequently one expects the asymptotic limit to be plane fronted waves with colliding wave fronts (the so--called Kundt \cite{Kundt:1961} waves) in the most general case. We use the term `spherical waves' here to denote waves possessing compact and expanding wave fronts. The asymptotic limit we will display here is a consequence of our recent geometrical construction of plane and spherical gravitational waves \cite{Hogan:Puetzfeld:2021:1} providing a scheme to determine whether or not the waves collide. We begin by utilizing it to describe spherical electromagnetic waves propagating in a vacuum and their asymptotic plane fronted limit. This is followed by constructing the asymptotic limit of Robinson--Trautman waves. The resulting limit is plane fronted waves which may or may not be colliding. We make use of a novel form of the Robinson--Trautman wave line element to achieve this result. The paper ends with a brief summary and an interesting observation. In appendix \ref{app:1} we provide details on the derivation of the novel form of the Robinson--Trautman wave line element.

\section{Spherical Electromagnetic Waves}\label{sec:1}

We consider spherical electromagnetic waves in the context of Minkowskian spacetime in which the histories of the wave fronts are null cones with vertices on an arbitrary world line. A novel form of the Minkowskian line element tailor--made for our purposes is given by \cite{Hogan:Puetzfeld:2021:1}
\begin{equation}\label{1}
ds_0^2=2\left (1-\frac{m\,v}{2}\right )^2\left |d\zeta-\frac{v\,q_{\bar\zeta}}{\left (1-\frac{m\,v}{2}\right )}du\right |^2+2\,q\,du\,dv\ ,
\end{equation}
with $m(u)$ an arbitrary real--valued function of $u$, $\zeta$ a complex coordinate with complex conjugate $\bar\zeta$  and
\begin{equation}\label{2}
q(\zeta, \bar\zeta, u)=\bar\beta(u)\,\zeta+\beta(u)\,\bar\zeta+\frac{1}{2}\alpha(u)\,\zeta\,\bar\zeta+\frac{1}{2}\gamma(u)\ ,\ \ q_{\bar\zeta}=\frac{\partial q}{\partial\bar\zeta}\ ,
\end{equation}
where $\beta(u)$ is an arbitrary complex--valued function of $u$, $\alpha(u)=dm(u)/du$ and $\gamma(u)$ is an arbitrary real--valued function of $u$. Note that it was shown in \cite{Hogan:Puetzfeld:2021:1} that Kundt and pp-waves can be described in a unified way, in which the function $\beta$ controls the branch of the solution of these formerly distinct line elements. In coordinates $x^{i'}=(\zeta, \bar\zeta, u, v)$ the hypersurfaces $u={\rm constant}$ are null cones whose generators are labelled with the complex coordinate $\zeta$ and the real coordinate $v$ is an affine parameter along them with $v=2/m(u)$ corresponding to the vertex of each null cone. The vector field $\partial/\partial v$ is null, geodesic, shear--free,twist--free and has expansion $-\frac{m}{2}\left (1-\frac{m\,v}{2}\right )^{-1}$. A convenient \emph{distance} from the world line of the vertices of the null cones is the \emph{parallax distance} \cite{Sachs:1962} $r$ given by 
\begin{equation}\label{2'}
r=\frac{2}{m}-v\ .
\end{equation}
Since $r\rightarrow+\infty$ corresponds to $m\rightarrow 0$ the asymptotic limit which interests us below will be achieved by taking the limit $m\rightarrow 0$.

The relationship between the coordinates $x^{i'}$ and rectangular Cartesians and time $X^i=(X, Y, Z, T)$ resulting in
\begin{equation}\label{3}
ds_0^2=(dX)^2+(dY)^2+(dZ)^2-(dT)^2=\eta_{ij}\,dX^i\,dX^j\ ,
\end{equation}
is given by
\begin{eqnarray}
\zeta&=&\left (1-\frac{m}{2}(Z-T)\right )^{-1}\left\{\frac{1}{\sqrt{2}}(X+i\,Y)+l(u)(Z-T)\right\}\ ,\nonumber \\
\label{4}\\
v&=&Z-T\ ,\label{5}
\end{eqnarray}
and $u(X, Y, Z, T)$ given implicitly by
\begin{eqnarray}
Z&+&T=\frac{m}{2}\left (1-\frac{m}{2}(Z-T)\right )^{-1}
\nonumber\\
&&\times|X+i\,Y+\sqrt{2}\,l(u)(Z-T)|^2+\sqrt{2}\,l(u)\,(X-i\,Y)\nonumber\\
&&+\sqrt{2}\,\bar l(u)\,(X+i\,Y)+2\,l(u)\,\bar l(u)\,(Z-T)+n(u)\ . \nonumber \\
\label{6}
\end{eqnarray}
The geometrical origin of these equations can be found in \cite{Hogan:Puetzfeld:2021:1}. A similar geometrical construction in the presence of a cosmological constant has been described in \cite{Hogan:2018:1}. The partial derivatives of $\zeta, v, u$ with respect to the rectangular Cartesians and time $X^i$ are indicated by a comma and given via the 1--forms:
\begin{eqnarray}
d\zeta&=&\zeta_{,i}\,dX^i\nonumber\\
&=&\left (1-\frac{m}{2}(Z-T)\right )^{-1}\Biggl\{\frac{1}{\sqrt{2}}(dX+i\,dY)\nonumber\\
&&+\left(l(u)+\frac{1}{2}m(u)\,\zeta\right )(dZ-dT)\nonumber\\
&&+\left (\beta(u)+\frac{1}{2}\alpha(u)\,\zeta\right )\,(Z-T)du\Biggr\}\ ,\label{7}\\
dv&=&v_{,i}\,dX^i=dZ-dT\ ,\label{8}\\
du&=&u_{,i}\,dX^i=-\frac{m(u)}{2\,q}\left (1-\frac{m(u)}{2}(Z-T)\right )^{-1}\xi_i\,dX^i\ ,\nonumber\\
&=&-\frac{m(u)}{2\,q}\left (1-\frac{m(u)}{2}(Z-T)\right )^{-1}\nonumber\\
&&\Biggl\{\left (X+\frac{\sqrt{2}}{m(u)}(\bar l(u)+l(u))\right )\,dX\nonumber\\
&&+\left (Y+\frac{i\sqrt{2}}{m(u)}(\bar l(u)+l(u))\right )dY \nonumber\\
&&+\left (Z-\frac{n(u)}{2}+\frac{2\,l(u)\bar l(u)-1}{m(u)}\right )dZ\nonumber\\
&&-\left (T-\frac{n(u)}{2}+\frac{2\,l(u)\bar l(u)+1}{m(u)}\right )dT\Biggr\}\ .\label{9}
\end{eqnarray}
The final 1--form defines the 4--vector $\xi^i$ with $\xi_i=\eta_{ij}\,\xi^j$ and since $u_{,i}$ is a null vector field we have $\eta_{ij}\,\xi^i\,\xi^j=0$. Also $q(\zeta, \bar\zeta, u)$ is given by (\ref{2}) with $\beta(u)=dl(u)/du$ and $\gamma(u)=dn(u)/du$. 

To describe spherical fronted electromagnetic waves propagating in a vacuum we use a potential 1--form 
\begin{equation}\label{10}
A=f(\zeta, \bar\zeta, u)\,du\ ,
\end{equation}
with $f(\zeta, \bar\zeta, u)$ a real valued function of its argument. The Maxwell field is given by the 2--form
\begin{equation}\label{11}
F=dA=\frac{\partial f}{\partial\zeta}\,d\zeta\wedge du+\frac{\partial f}{\partial\bar\zeta}\,d\bar\zeta\wedge du\ ,
\end{equation}
which satisfies Maxwell's vacuum field equations provided
\begin{equation}\label{12}
\frac{\partial^2f}{\partial\zeta\partial\bar\zeta}=0\ .
\end{equation}
Hence $\partial f/\partial\zeta=g(\zeta, u)$ is an arbitrary complex--valued analytic function. It is convenient to work with the complex 2--form
\begin{equation}\label{13}
{\cal F}=F-i\,{}^*F=2\,g(\zeta, u)\,d\zeta\wedge du\ \ \ {\rm with}\ \ {}^*{\cal F}=i\,{\cal F}\ ,
\end{equation}
and the star denotes the Hodge dual. Before taking the asymptotic limit we express this Maxwell field in terms of the coordinates $X^i=(X, Y, Z, T)$, using (\ref{7}) and (\ref{9}), and read off the electric 3--vector ${\bf E}$ and the magnetic 3--vector ${\bf B}$. In this way we arrive at the vectors ${\bf E}, {\bf B}$ given in Bateman form \cite{Bateman:1955:1} in terms of Jacobian determinants by
\begin{eqnarray}
B^1+iE^1&=&-2\,g(\zeta, u)\,\frac{\partial(\zeta, u)}{\partial(Y, Z)}\ ,\label{s9}\\
B^2+iE^2&=&-2\,g(\zeta, u)\,\frac{\partial(\zeta, u)}{\partial(Z, X)}\ ,\label{s10}\\
B^3+iE^3&=&-2\,g(\zeta, u)\,\frac{\partial(\zeta, u)}{\partial(X, Y)}\ ,\label{s11}
\end{eqnarray}
with the second of (\ref{13}) requiring
\begin{eqnarray}
\frac{\partial(\zeta, u)}{\partial(Z, X)}&=&i\,\frac{\partial(\zeta, u)}{\partial(Y, T)}\nonumber\\
&=&\frac{\sqrt{2}\left\{(\xi^1+i\,\xi^2)\xi^1+(\xi^3-\xi^4)\xi^3\right\}}{q\,m\,(\xi^3-\xi^4)^3}\ ,\label{s12}\\
\frac{\partial(\zeta, u)}{\partial(Y, Z)}&=&i\,\frac{\partial(\zeta, u)}{\partial(X, T)}\nonumber\\
&=&-\frac{\sqrt{2}\left\{(\xi^1+i\,\xi^2)\xi^2+i(\xi^3-\xi^4)\xi^3\right\}}{q\,m\,(\xi^3-\xi^4)^3}\ ,\label{s13}\\
\frac{\partial(\zeta, u)}{\partial(X, Y)}&=&i\,\frac{\partial(\zeta, u)}{\partial(Z, T)}=\frac{i\sqrt{2}\,(\xi^1+i\,\xi^2)}{q\,m\,(\xi^3-\xi^4)^2}\ ,\label{s14}
\end{eqnarray}
with $q$ given by (\ref{2}) and $\xi^i=\eta^{ij}\,\xi_j$ (with $\eta^{ij}\,\eta_{jk}=\delta^i_k$) given by (\ref{9}). We can write (\ref{s9})--(\ref{s11}) neatly in the form
\begin{equation}\label{s15}
{\bf B}+i{\bf E}=-\frac{4\,i\,g(\zeta, u)\,\xi^4}{q\,m\,(\xi^3-\xi^4)^2}\,{\bf M}\ ,
\end{equation}
with the complex 3--vector ${\bf M}=(M^1, M^2, M^3)$ given by
\begin{eqnarray}
M^1&=&\frac{i(\xi^1+i\xi^2)\xi^2-(\xi^3-\xi^4)\xi^3}{\sqrt{2}\,\xi^4(\xi^3-\xi^4)}\ ,\label{s16}\\
M^2&=&\frac{-i(\xi^1+i\xi^2)\xi^1-i(\xi^3-\xi^4)\xi^3}{\sqrt{2}\,\xi^4(\xi^3-\xi^4)}\ ,\label{s17}\\
M^3&=&\frac{(\xi^1+i\xi^2)}{\sqrt{2}\,\xi^4}\ .\label{s18}
\end{eqnarray}
Furthermore defining the complex variable
\begin{equation}\label{14}
L(\zeta, u)=\frac{1}{\sqrt{2}}\left (\frac{\xi^1+i\,\xi^2}{\xi^4-\xi^3}\right )=l(u)+\frac{1}{2}m(u)\,\zeta\ ,
\end{equation}
we can write ${\bf M}$ in the form
\begin{equation}\label{15}
{\bf M}=\left (\frac{1-2\,L^2}{\sqrt{2}\,(2\,L\,\bar L+1)}\ ,\ \frac{i(1+2\,L^2)}{\sqrt{2}\,(2\,L\,\bar L+1)}\ ,\ \frac{2\,L}{2\,L\,\bar L+1}\right )\ ,
\end{equation}
and (\ref{s15}) can now be written as
\begin{equation}\label{16}
{\bf B}+i\,{\bf E}=-i\,g(\zeta, u)\,q^{-1}\left (1-\frac{m\,v}{2}\right )^{-1}(2\,L\,\bar L+1)\,{\bf M}\ .
\end{equation}
The histories in Minkowskian space--time of the spherical wave fronts are the null cones $u(X^i)={\rm constant}$ and so the \emph{wave velocity} has components \cite{Synge:1965}
\begin{equation}\label{s20}
v^{\alpha}=-\frac{u_{,4}\,u_{,\alpha}}{u_{,\beta}\,u_{,\beta}}=\frac{\xi^{\alpha}}{\xi^4}\ \ \Rightarrow {\bf v}\cdot{\bf v}=1\ ,
\end{equation}
with Greek indices taking values 1, 2, 3. In terms of $L$ in (\ref{14}) we have
\begin{equation}\label{s20'}
{\bf v}=\left (\frac{\sqrt{2}\,(\bar L+L)}{2\,L\,\bar L+1}, \frac{i\sqrt{2}\,(\bar L-L)}{2\,L\,\bar L+1}, \frac{2\,L\,\bar L-1}{2\,L\,\bar L+1}\right )\ .
\end{equation}
Denoting the complex conjugate of ${\bf M}$ by $\bar{\bf M}$ we find that ${\bf M}, \bar{\bf M}, {\bf v}$ satisfy the algebraic conditions:
\begin{equation}\label{s21}
{\bf M}\cdot{\bf M}=0\ ,\ {\bf M}\cdot\bar{\bf M}=1\ \ {\rm and}\ \ {\bf M}\times\bar{\bf M}=i\,{\bf v}\ .
\end{equation}
It thus follows from (\ref{16}) that (a) $|{\bf E}|=|{\bf B}|$ and ${\bf E}\cdot{\bf B}=0$ (the algebraic conditions for pure electromagnetic radiation) , (b) ${\bf E}, {\bf B}, {\bf v}$ constitute a right handed triad and (c) the Poynting vector ${\bf E}\times{\bf B}=|{\bf E}|^2{\bf v}$ (so that the energy of the waves is propagated with velocity ${\bf v}$). The question of whether or not the wave fronts collide depends crucially on the behavior of the world line of the vertices of the histories of the wave fronts (which follows from the detailed discussion in \cite{Hogan:Puetzfeld:2021:1}). This world line is the history in Minkowskian space-time of the point--like source of the electromagnetic waves. The asymptotic limit of (\ref{1}) and (\ref{16}) is obtained taking the limit $m\rightarrow 0$ resulting in
\begin{equation}\label{17}
\lim_{m\rightarrow 0}ds_0^2=2\,|d\zeta-v\,q_{\bar\zeta}\,du|^2+2\,q\,du\,dv\ ,
\end{equation}
and
\begin{equation}\label{18}
\lim_{m\rightarrow 0}({\bf B}+i\,{\bf E})=-i\,g(\zeta, u)q^{-1}(2\,l(u)\bar l(u)+1)\,{\bf m}\ ,
\end{equation}
with now
\begin{equation}\label{19}
q=\beta(u)\,\bar\zeta+\bar\beta(u)\,\zeta+\frac{1}{2}\gamma(u)\ ,
\end{equation}
and
\begin{equation}\label{20}
{\bf m}= \left ( \frac{1-2\,l^2}{\sqrt{2}\,(2\,l\,\bar l+1)}, \frac{i(1+2\,l^2)}{\sqrt{2}\,(2\,l\,\bar l+1)}, \frac{2\,l}{2\,l\,\bar l+1}\right )\ .
\end{equation}
Here $l=\lim_{m\rightarrow 0}L$ following from (\ref{14}). These are plane fronted electromagnetic waves with histories $u={\rm constant}$ which are null hyper\emph{planes} which intersect so long as $\beta(u)\neq0$ (see \cite{Hogan:Puetzfeld:2021:1}).

\section{Spherical Gravitational Waves}\label{sec:2}

The spherical gravitational waves of interest to us are described by the Robinson--Trautman \cite{Robinson:Trautman:1960,Robinson:Trautman:1962} purely radiative solutions of Einstein's vacuum field equations. In the coordinates $x^i=(\zeta, \bar\zeta, u, v)$ we are using we have a novel form of the Robinson--Trautman solutions of this type given by the line element
\begin{eqnarray}
ds^2&=&2\,\left (1-\frac{m\,v}{2}\right )^2\left |d\zeta+\left (G-\frac{v\,q_{\bar\zeta}}{1-\frac{m\,v}{2}}\right )du\,
\right |^2\nonumber\\
&+&2\,q\,du\,\Biggl\{dv+\frac{1}{m}\left (1-\frac{m\,v}{2}\right )(G_{\zeta}+\bar G_{\bar\zeta})du\Biggr\}. \label{21}
\end{eqnarray}
Here $G(\zeta, u)$ is an arbitrary analytic function with $G_{\zeta}=\partial G/\partial\zeta$ and the bar, as always, denoting complex conjugation. Also $q$ and $q_{\zeta}$ are as in (\ref{2}). This form of line element is derived in the appendix \ref{app:1} from the standard Robinson--Trautman form using the equations given in section III of \cite{Hogan:Puetzfeld:2021:1}. Einstein's vacuum field equations are satisfied by the metric given by this line element. The Newman--Penrose \cite{Newman:Penrose:1962} components $\Psi_{A}$ with $A=0, 1, 2, 3, 4,$ of the Riemann curvature tensor vanish with the exception of 
\begin{equation}\label{22}
\Psi_4=(m\,q)^{-1}\left (1-\frac{m\,v}{2}\right )^{-1}\frac{\partial^3G}{\partial\zeta^3}\ ,
\end{equation}
which corresponds to a purely radiative curvature tensor in the Petrov classification with degenerate principal null direction, and propagation direction of the history of the radiation in spacetime, given by the vector field $\partial/\partial v$. Putting
\begin{equation}\label{23}
G(\zeta, u)=m(u)\,{\cal G}(\zeta, u)\ ,
\end{equation}
and taking the asymptotic limit of (\ref{21}) and (\ref{22}), which corresponds to the limit $m\rightarrow 0$, we clearly arrive at
\begin{equation}\label{24}
\lim_{m\rightarrow 0}ds^2=2\,|d\zeta-v\,q_{\bar\zeta}\,du|^2+2\,q\,du\,\{dv+({\cal G}_{\zeta}+\bar{\cal G}_{\bar\zeta})du\}\ ,
\end{equation}
with now
\begin{equation}\label{25}
q(\zeta, \bar\zeta, u)=\bar\beta(u)\,\zeta+\beta(u)\,\bar\zeta+\frac{1}{2}\gamma(u)\ .
\end{equation}
In addition we have from (\ref{22})
\begin{equation}\label{26}
\lim_{m\rightarrow 0}\Psi_4=q^{-1}\frac{\partial^3{\cal G}}{\partial\zeta^3}\ .
\end{equation}
In (\ref{24})--(\ref{26}) we have arrived at plane fronted waves with colliding wave fronts (Kundt \cite{Kundt:1961} waves) if $\beta(u)\neq 0$ or plane fronted waves with noncolliding wave fronts (pp--waves \cite{Brinkmann:1923}) if $\beta(u)=0$. This limiting form of the line element and curvature tensor incorporating both of these special cases has been derived geometrically in \cite{Hogan:Puetzfeld:2021:1}.

We have demonstrated explicitly in \cite{Hogan:Puetzfeld:2021:1} that $\beta(u)\neq0$ means that the normals to the null hyperplane histories of the wave fronts are not covariantly constant and that $\beta(u)\neq0$ corresponds to intersecting null hyperplanes and therefore to colliding wave fronts.

\section{Discussion}\label{sec:3}

We have demonstrated explicitly that plane fronted electromagnetic or gravitational waves undergo collisions in general since they are the asymptotic limit of spherical fronted waves which have isolated sources moving arbitrarily. The role of the function $q$ on the right hand side of (\ref{18}) in the case of the electromagnetic field of plane fronted electromagnetic waves and on the right hand side of (\ref{26}) in the case of the gravitational field of plane fronted gravitational waves is interesting. If the waves are simple plane waves then the right hand sides of (\ref{18}) and (\ref{26}) must be functions of $u$ only. This can only occur if $q$ is independent of $\zeta$ and $\bar\zeta$ so that $\beta(u)=0$. It follows therefore from \cite{Hogan:Puetzfeld:2021:1} that simple plane waves of this kind cannot collide.

\begin{acknowledgments}
This work was funded by the Deutsche Forschungsgemeinschaft (DFG, German Research Foundation) through the grant PU 461/1-2 -- project number 369402949 (D.P.). 
\end{acknowledgments}

\appendix

\section{Novel Form of the Robinson--Trautman Wave Line Element}\label{app:1}

The classical form of the Robinson--Trautman line element in the special case of a vacuum field containing pure gravitational waves is given by \cite{Hogan:Puetzfeld:2021:1}
\begin{equation}\label{A1}
ds^2=2\,r^2\,p^{-2}|d\zeta+Q\,du|^2-2\,du\,dr-(K-2\,H\,r)\,du^2\ ,
\end{equation}
with
\begin{equation}\label{A2}
p=\frac{2\,q}{m}\ ,\ Q=\frac{2\,q_{\bar\zeta}}{m}+G(\zeta, u)\ ,\ K=-\frac{\kappa}{m^2}\ ,
\end{equation}
and
\begin{equation}\label{A3}
H=q^{-1}\dot q-\frac{4}{m}q^{-1}q_{\zeta}\,q_{\bar\zeta}-q^{-1}\,q_{\zeta}\,G-q^{-1}\,q_{\bar\zeta}\,\bar G+\frac{1}{2}(G_{\zeta}+\bar G_{\bar\zeta})\ .
\end{equation}
Here $q(\zeta, \bar\zeta, u)$ and $m(u)$ are as they appear in equations (\ref{1}) and (\ref{2}), $G(\zeta, u)$ is an arbitrary analytic function, and
\begin{equation}\label{A4}
\kappa=2\,|\beta|^2-\frac{1}{2}\,\alpha\,\gamma\ ,
\end{equation}
with $\alpha(u)$, $\beta(u)$ and $\gamma(u)$ as they appear in equation (\ref{2}). Now introduce a coordinate $v$ by writing
\begin{equation}\label{A5}
r=\frac{2\,q}{m}\,\left (1-\frac{m\,v}{2}\right )\ .
\end{equation}
It thus follows that
\begin{eqnarray}
dr&=&-q\,dv+\frac{2}{m}\left (1-\frac{m\,v}{2}\right )\,(q_{\zeta}\,d\zeta+q_{\bar\zeta}\,d\bar\zeta) \nonumber \\
&&+2\,\left\{\frac{\dot q}{m}\left (1-\frac{m\,v}{2}\right )-\frac{q\,\alpha}{m^2}\right\}\,du\ ,\label{A6}
\end{eqnarray}
with the dot indicating partial differentiation with respect to $u$ and remembering that $\alpha=\dot m$. With this we can write (\ref{A1}) in the form
\begin{eqnarray}
ds^2&=&2\,\left (1-\frac{m\,v}{2}\right )^2d\zeta\,d\bar\zeta+2\,q\,du\,dv+C\,du^2\  \nonumber \\
&&+2\,\left (1-\frac{m\,v}{2}\right )^2\left\{\bar G-\frac{v\,q_{\zeta}}{\left (1-\frac{m\,v}{2}\right )}\right\}\,du\,d\zeta\nonumber\\
&&+2\,\left (1-\frac{m\,v}{2}\right )^2\left\{G-\frac{v\,q_{\bar\zeta}}{\left (1-\frac{m\,v}{2}\right )}\right\}\,du\,d\bar\zeta,\label{A7}
\end{eqnarray}
with
\begin{eqnarray}
C&=&\frac{2\,q}{m}\left (1-\frac{m\,v}{2}\right )\,(G_{\zeta}+\bar G_{\bar\zeta})+2\,\left (1-\frac{m\,v}{2}\right )^2G\,\bar G\nonumber \\
&&-2\,v\,q_{\bar\zeta}\,\left (1-\frac{m\,v}{2}\right )\,\bar G-2\,v\,q_{\zeta}\,\left (1-\frac{m\,v}{2}\right )\,G\nonumber\\
&&+\frac{4\,q\,\alpha}{m^2}+\frac{4\,\kappa}{m^2}-\frac{8\,q_{\zeta}\,q_{\bar\zeta}}{m^2}\left (1-\frac{m^2v^2}{4}\right )\ .\label{A8}
\end{eqnarray}
But
\begin{equation}\label{A9}
q_{\zeta}\,q_{\bar\zeta}=\frac{1}{2}(\kappa+\alpha\,q)\ ,
\end{equation}
and so we can write
\begin{eqnarray}
C&=&\frac{2\,q}{m}\,\left (1-\frac{m\,v}{2}\right )(G_{\zeta}+\bar G_{\bar\zeta})\nonumber \\
&&+2\,\left (1-\frac{m\,v}{2}\right )^2 \left |G-\frac{v\,q_{\bar\zeta}}{\left (1-\frac{m\,v}{2}\right )}\right |^2\ .\label{A10}
\end{eqnarray}
When this is substituted into (\ref{A7}) the line element (\ref{21}) results.

\bibliographystyle{unsrtnat}
\bibliography{planelimit_bibliography}
\end{document}